\documentclass[amsmath,amssymb,aps,nofootinbib]{revtex4}
\usepackage[dvips]{graphicx}
\usepackage{graphicx}
\usepackage{amsfonts}
\usepackage{bm}
\usepackage{amsmath}
\usepackage{amssymb}
\usepackage{color}
\usepackage[all]{xy}

\def\be{\begin{equation}}
\def\ee{\end{equation}}
\def\bea{\begin{eqnarray}}
\def\eea{\end{eqnarray}}

\newcommand{\rd}{\mathrm{d}} 
\newcommand{\ri}{\mathrm{i}} 
\def\tr{
{\rm tr}}

\begin{document}

\title{Topological Quantum Numbers of Dyonic Fields over Taub-NUT and Taub-Bolt Spaces}
\author{Daniel Flores-Alfonso$^1$ and Hernando Quevedo$^{1,2}$}
\email{daniel.flores@correo.nucleares.unam.mx,quevedo@nucleares.unam.mx}
\affiliation{$^1$Instituto de Ciencias Nucleares,
Universidad Nacional Aut\'onoma de M\'exico,\\
 AP 70543, Ciudad de M\'exico 04510, M\'exico \\
$^2$Department of Theoretical and Nuclear Physics,
 Kazakh National University, 
Almaty 050040, Kazakhstan}

\begin{abstract}
We calculate the Chern numbers of ${\rm SU}(2)$-homogeneous Einstein-Maxwell gravitational instantons
with boundary at infinity. By restating these numbers as Chern-Simons invariants on the boundary 
apparent conflicting results emerge. We resolve this issue examining the topological stability of the
self-gravitating Abelian fields. No quantization carrying physical meaning is found when the background
is a Taub-NUT space. However the magnetic charge of dyons on Taub-Bolt spaces is found to be of
topological quantum nature. In this framework electric charge is quantized by a consistency condition.

{MSC}: 
83C45, 53Z05

{Keywords}:
Topological Charge, Instantons, Taub-Bolt, Dyons

\end{abstract}

\maketitle

\label{first}

\section{Introduction}

The study of gravitational instantons in Euclidean Einstein gravity was initiated in the 70's \cite{ginstantons,bel,action,multi,eh,egh}  to apply the path-integral approach in the 
hope to construct a  theory of quantum gravity.  This expectation was based upon the success reached by this approach in the application of quantum field 
theory to Yang-Mills theories. Indeed, the instanton solutions of classical Euclidean Yang-Mills field equations, with self-dual field strength, allow one 
to interpret the physical vacuum state as a superposition of an infinite number of vacuum states. This leads to an alternative 
non-perturbative quantization of Yang-Mills theories by using the path-integral approach.  The idea of applying similar analysis to Euclidean 
Einstein gravity found several conceptual and technical difficulties \cite{carlip}, which are currently under investigation 
to formulate a consistent theory of quantum gravity. 
Nevertheless, gravitational instantons became a topic of intensive research, and together with monopoles and solitons constitute the area of 
theoretical physics which today is  known as topological defects. 

A different approach was recently proposed in which  concepts of topological quantization are applied to find quantum information from 
classical fields, including gravitational configurations \cite{leo}. In this case, the underlying classical structure is different. 
To any classical configuration, which includes gauge fields, gravitational fields and mechanical systems, one associates a principal fiber bundle 
where the base space should contain all the information of the classical configuration and the standard fiber corresponds to its physical symmetries. 
In the case of gravity, for instance, the base space is the spacetime and the standard fiber is the Lorentz group 
${\rm SO}(3,1)$ so that the bundle is 10-dimensional.  The main goal of topological quantization is to show that the topological and geometric properties 
of the principal fiber bundle can be used to obtain quantum information about the corresponding classical configuration. 
This goal has been reached only partially \cite{tqclass,tqmech,net,gus,gus2}. In particular, the topological numbers of the bundle are used to construct a topological 
spectrum whose physical information is directly related to the information contained in the spectrum obtained in canonical quantization. 
Topological invariants are therefore an important tool of topological quantization and this is one of the reasons why we are interested 
in analyzing the properties of topological defects.
      
Recently, the classic subject of topological defects \cite{dirac,thooft,polyakov,bpst,wu} has been applied to condensed 
matter, i.e., topological insulators --- for a review on the subject see, for example, Ref. \cite{notes}. The main r\^ole is played by topological quantum numbers
such as Chern invariants.
The subject includes other aspects of topology as well. These subjects are related physically to
topologically protected states, time-reversal symmetry, topological phases of matter and topological phase transitions (for a 
a selection of recent references see  \cite{42,31,32,50,47,39,49,35,24} and
earlier works in \cite{60,59,17,13,15}). The importance of this research was notoriously recognized in the 2016 Nobel prize which was awarded to D.J. Thouless, F.D.M. Haldane and J.M. Kosterlitz ``for theoretical 
discoveries of topological phase transitions and topological phases of matter''.

Phase transitions which include a change in topology have been studied in gravitation as well.
Research in this direction usually entails using gravitational instantons --- the gravitational analogs
of Yang-Mills pseudo-particles\cite{ginstantons}.
These solutions have positive definite metric. Euclidean metrics may be hard to be interpreted physically,
but are a useful tool for calculating the free energy of the corresponding Lorentzian system. 
For example, in the presence of a negative cosmological constant, 
radiation over the AdS space and the AdS-Schwarzschild black hole are phases of the same thermodynamic system. 
The partition function may be estimated
using their Wick rotated counterparts. Topology changes during these phase transitions. Thermodynamics \cite{hp} and Ricci flow with 
surgery \cite{ricciflow0} 
are used to grasp these changes. These transitions have been studied for Taub-NUT and Taub-Bolt spacetimes as well \cite{cvj,ricciflow}.

It is very well known \cite{eh,sen} that a self-dual Maxwell field naturally lives on the Taub-NUT
instanton. This field is dyonic. It is a renowned \cite{ginstantons} result that the Taub-NUT instanton is topologically trivial.
One usually expects that the Chern number of such spaces vanish. However, this is not the case for Taub-Bolt.
In this work, we study dyonic Maxwell fields over Taub-Bolt which are not self-dual.
We focus on electromagnetic charge, as a topological quantum number, on a Taub-Bolt space --- which 
is a $4-$dimensional gravitational instanton. 

Topological quantum numbers originate with the Dirac monopole which has magnetic charge $p$ and gauge potential
\begin{equation}
 A=p \cos\theta \rd\phi   \ . \label{dirac}
\end{equation}
Since we are mainly concerned with gravitation in this paper, we review its unique self-gravitating
equivalent, i.e., the  magnetic Reissner-Nordstr\"om spacetime. The gauge potential is still given by Eq.(\ref{dirac}), 
but is defined over the space
$\mathbb{R}^2\times \mathbb{S}^2$, the topology of the Reissner-Nordstr\"om black hole. In spherical coordinates, $A$
is singular at $\theta=0$ and $\theta=2\pi$. To resolve this issue, we consider two gauge potentials equivalent to Eq. (\ref{dirac})
each defined on complementary open covers of the spheres coordinated by $(\theta, \phi)$.
\begin{equation}
 A_{\pm}=p(\pm1+\cos\theta)\rd\phi\ .
\end{equation}
Notice that $A_+-A_-=2p\rd\phi$. Consistency of the ${\rm U}(1)-$bundle where $A$ is determined requires
\begin{equation}
 2p=\text{integer} \label{int}
\end{equation}
meaning that the magnetic charge is quantized.
Moreover, the topology of the bundle space is then $\mathbb{S}^3/\mathbb{Z}_n$ whenever $p=n/2$ hence $p$ is called a topological charge.

Alternatively, we can  look at the Chern invariants of the configuration. 
Since our base manifold is four dimensional, there are two relevant Chern classes to consider, i.e.,
the two-form \cite{egh}
\begin{equation}
 c_1 =-\frac{1}{2\pi}\tr F 
\label{c1}
\end{equation}
and the four-form
\begin{equation}
 c_2 =\frac{1}{8\pi^2}(\tr F\wedge F-\tr F \wedge \tr F) \ .
\label{c2}
\end{equation}
This last cohomology class vanishes for ${\rm U}(1)-$bundles, the gauge group 
of electromagnetism, so we consider
only the two-form $c_1$. This Chern class corresponds to the field strength, defined by $F=\rd A$. 
The sphere representing the event horizon and all 2D submanifolds that can be deformed into it form a special homology class.
This leads to an integral formula
\begin{equation}
 \int\limits_{\mathbb{S}^2}c_1=2p
\end{equation}
Since the Chern class is an integer,  we obtain, once again, Eq.(\ref{int}).

Another question is what happens if the geometry and field are that of electric Reissner-Nordstr\"om. 
This question has been  tackled in \cite{leo}. Moreover, the generalization to include rotation has been considered as well. 
Indeed, the Schwarzschild spacetime may be generalized to Kerr-Newman, but
independently may be generalized to spaces with a NUT parameter.  
Specifically, we consider charged versions of Taub-NUT spaces.

This work is organized as follows. In Sec. \ref{nutsandbolts}, we review the main properties of the gravitational instantons that are 
generated by applying a Wick rotation to the  Taub-NUT spacetime with electric and magnetic charge. In Sec. \ref{sec:tqn}, we calculate 
explicitly the topological quantum numbers of dyonic fields on the Euclidean Taub-NUT and Taub-Bolt backgrounds. Finally, in Sec. \ref{sec:con}, 
we conclude with some remarks about possible future applications of our results.

\section{Nuts, Bolts and Dyons}
\label{nutsandbolts}
%
%
Some gravitational instantons may be obtained through Wick rotating a Lorentzian metric, this is the case of the Taub-NUT and 
Taub-Bolt\footnote{Taub-Bolt is the counter part of Taub-NUT which has a surface analog to an event horizon. See section \ref{nutsandbolts}.} instantons.
The Lorentzian Taub-NUT spacetime \cite{taub,nut} is defined over the space $\mathbb{S}^3\times\mathbb{R}$, and its metric is given by
\begin{equation}
\rd s^2 = -V(r)(\rd t+2L\cos\theta \rd\phi)^2+\frac{\rd r^2}{V(r)}+(r^2+L^2)(\rd\theta^2+\sin^2\theta \rd\phi^2) \label{ltn}
\end{equation}
with
\begin{equation}
 V(r)=\frac{r^2-2Mr-L^2}{r^2+L^2} \ .
\end{equation}

The underlying topology with periodic time $t\sim t+8\pi L$ is chosen to avoid the physical presence of Misner strings \cite{misner}.
The physical parameters are the (ordinary) gravitoelectric mass $M$ and the NUT charge $L$
also called the gravitomagnetic mass. The Taub-NUT metric represents a gravitational dyon, and  is asymptotically locally flat (ALF). 
Indeed, its curvature in the ${\rm SO}(3,\mathbb{C})-$representation \cite{solutions}, with a local orthonormal metric, can be expressed as \cite{quev92}
\begin{equation}
{\cal R} = \frac{M+\ri L}{(r+\ri L)^3} 
\left(\begin{array}{ccc}
-1 & & \\
   &-1 & \\
	& & 2
\end{array}\right)
\label{so3c}
\end{equation}
which is clearly locally flat, although the global topology at infinity can differ from that of the ordinary Euclidean space.

Taub-NUT metrics belong to the Bianchi IX class, i.e., spacetimes with an ${\rm SU}(2)$ homogeneity group.
Bianchi IX metrics have a portrayal as a sequence of squashed three-spheres.
Taub-NUT is a very symmetric case as its isometry group is ${\rm SU}(2)\times {\rm U}(1)$, where the ${\rm U}(1)$
subgroup on the right is a cyclic symmetry between two of the invariant directions of the spatial three-spheres. 
For this reason
Taub-NUT is called a biaxial Bianchi IX spacetime. 
So the sequence portraying the NUT space has
squashed circles Hopf fibered over round two-spheres.
This singles out one direction, the one in which circles are fibered over spheres.

A generalization of the Taub-NUT metric that includes Maxwell fields was found in \cite{brill}.
The line element is still written as in Eq.(\ref{ltn}), but the function $V(r)$ instead takes the form
\begin{equation}
V(r)= \frac{r^2-2Mr-L^2+Q^2+p^2}{r^2+L^2} \ .
\end{equation}
The construction of this metric takes advantage of the biaxial symmetry described above. The gauge potential is
\begin{equation}
 A=\left(\frac{Qr}{r^2+L^2}+\frac{p}{2L}\frac{r^2-L^2}{r^2+L^2}\right)(\rd t+2L\cos\theta \rd\phi)\label{dyon}\ .
\end{equation}
In the $L\rightarrow0$ limit, we recover the dyonic Reissner-Nordstr\"om spacetime. 
Through a simple gauge transformation $A\rightarrow A-p/2L~dt$ 
we have the more convenient expression
\begin{equation}
 A=p\cos\theta \rd\phi+\left(\frac{Qr-pL}{r^2+L^2}\right)(\rd t+2L\cos\theta \rd\phi) \ .
\end{equation}

As before, the gauge potential is singular whenever $\sin\theta=0$ as apparently the magnetic part of the field vanishes.
We resolve this in the classical manner defining
\begin{equation}
 A_{\pm}=p(\pm1+\cos\theta)\rd\phi+\left(\frac{Qr-pL}{r^2+L^2}\right)(\rd t+2L\cos\theta \rd\phi) 
\end{equation}
concluding that the value $2p$ is an integer. However, the spacetime topology prevents this magnetic charge to be a topological quantum number.

To obtain the gravitational instantons related to this spacetime, we make a Wick rotation $t\mapsto \ri\tau$ on the neutral Taub-NUT spacetime.
Later on, we consider the charged version. However, as is readily seen,
the metric is only a real Riemannian metric if we also apply the transformation $L\mapsto \ri N$. 
So two geometrically distinct gravitational instantons arise.
Additionally, conical singularities emerge unless we impose certain restrictions which are detailed in the sequel.
The two Euclidean metrics have the form
\begin{equation}
 \rd s^2= V(r)(\rd\tau+2N\cos\theta \rd\phi)^2+\frac{\rd r^2}{V(r)}+(r^2-N^2)(\rd\theta^2+\sin^2\theta \rd\phi^2) \ .
\end{equation}
The $\mathbb{S}^3$ submanifold degenerates at $r=r_+$ where $V$ is nil --- the nature of this degeneration distinguishes the instanton pair.
Historically, the first case found was the self-dual Taub-NUT instanton \cite{ginstantons}, so called because its curvature form is self-dual. We henceforth
refer to this instanton systematically as Taub-NUT. In what follows Lorentzian spaces are no longer considered so no ambiguity arises.
The second case is called  Taub-Bolt  \cite{new} and was first written down in  \cite{bolt}.
The Taub-Bolt and Taub-NUT manifolds have different allowed ranges for $M$, $N$ and $r$ --- the appropriate conditions have been worked out 
in \cite{tnb}.

For Taub-NUT, the three-sphere degenerates into a point, called a nut, and $V$ reduces to 
\begin{equation}
 V_N=\frac{r-N}{r+N}\ .
\end{equation}
In this coordinate system there is a coordinate singularity at $r=N$ (called a nut). The global structure is that of the space $\mathbb{R}^4$.
So the Euler number and Hirzebruch signature of Taub-NUT are $\chi=1$ and $\sigma=0$, respectively.

In the Taub-Bolt case, the three-sphere degenerates into a two-sphere, via the Hopf fibration, which is called a bolt. In this case,
$V=V_B$ which is
\begin{equation}
V_B=\frac{(r-N/2)(r-2N)}{r^2-N^2}\ .
\end{equation}
As in the NUT case, the bolt (which is at $r=2N$) is a regular set, but in this coordinate system there is a coordinate singularity. 
Here $r=N/2$ is not contained within the range of $r$.
Moreover, the minimum value of $r$, $r_+$, 
is always greater than $N$, a property which is also valid in the generalizations of Taub-Bolt we treat later on.
The natural topology is that of the space $\mathbb{CP}^2\char`\\\{*\}$, which has $\chi=2$ and $\sigma=1$.

Both Taub-NUT and Taub-Bolt are limits of compact manifolds with an $\mathbb{S}^3$ boundary. The limit involves taking the boundary to infinity.
The bolt in the latter case is a minimal surface very similar to a black hole horizon. Moreover,
the space $\mathbb{CP}^2\char`\\\{*\}$ has the homotopy type
of $\mathbb{S}^2$ which happens in some black holes as well, e.g., Schwarzschild. 
The Taub-Bolt space has the topology of a cone Hopf fibered over $\mathbb{S}^2$ with the tip of the cone lying on the bolt. 
Since the cone is retractable, it is straightforward to see that the Taub-Bolt space indeed has the homotopy type of a sphere.
Physically, one can think of Taub-Bolt as an excitation of Taub-NUT, just as we can do
for Schwarzschild and flat space \cite{new}.

%
%
As expected, these vacuum solutions may be generalized to gravitational instantons of Euclidean Einstein-Maxwell theory by Wick 
rotating the dyon potential in Eq. (\ref{dyon}). To obtain a real gauge potential, we must take $Q\mapsto \ri q$. Notice that
the quantities $L$ and $Q$ are related to timelike directions in the Lorentzian configuration so Wick rotations affect them.
This happens for the angular momentum in rotating configurations as well. Since the parameter $p$ is magnetic, it remains unaffected.

The line element remains with the same structure, but the metric function $V$ instead takes the form
\begin{equation}
 V_d=\frac{r^2-2Mr+N^2+p^2-q^2}{r^2-N^2} 
\end{equation}
and the dyon potential is
\begin{equation}
 A=h(r)(\rd\tau+2N\cos\theta \rd\phi)\label{rdyon} 
\end{equation}
where
\begin{equation}
h(r)=\frac{qr}{r^2-N^2}+\frac{p}{2N}\frac{r^2+N^2}{r^2-N^2} \ .
\label{h}
\end{equation}
For these dyonic configurations, consistency again restricts the ranges of the physical parameters, including the electric charge. 
Given that we require $V_d(r_+)=0$ and $h(r_+)=0$, then mass and electric charge must comply with
\begin{equation}
 M=\frac{r_+^2+N^2+p^2-q^2}{2r_+}
\end{equation}
and
\begin{equation}
 q=-p\frac{r_+^2+N^2}{2Nr_+}\ .
\end{equation}

Since the NUT and Bolt geometry have the same asymptote, the charge and potential at infinity are equal for both cases.
The asymptotic charge and potential are $q$ and $p/2N$, respectively.

Taub-NUT and Taub-Bolt configurations have been also studied in generalized theories of gravity \cite{hendi06,hendi08}.

\section{Topological Quantum Numbers}
\label{sec:tqn}

Topological quantum numbers are at the heart of the study of topological defects. They represent the quantized conductance in the
quantum Hall effect\cite{60}. They describe the number of levels in molecular spectra\cite{molecular,molecular1}. They are the numbers characterizing monopole and
instanton solutions\cite{dirac,thooft,polyakov,bpst,wu}. In Skyrmions they represent the baryon number\cite{skyrme}. In $\mathbb{Z}_2$ topological insulators 
they distinguish between an ordinary insulator state and a quantum Hall state\cite{42}.

As a first example, let us consider again the magnetic Reissner-Nordstr\"om configuration. The background is a 4D manifold $M$ and has a special 2D submanifold, 
the two-sphere representing the event horizon\footnote{There are infinitely many equivalent special submanifolds, for example the two-sphere at infinity}. 
The quantity
 \begin{equation}
  C_1=c_1[\mathbb{S}^2]=2p=n
 \end{equation}
is an integer but, it is not a Chern number for $M$. Chern numbers are obtained from maximally dimensional forms in the Chern classes.

For an arbitrary unitary group bundle over a $4-$dimensional manifold $M$, the  Chern numbers are defined by \cite{egh}
\begin{equation}
 C_1^2=c_1\wedge c_1[M]\ , \qquad C_2 =c_2[M] 
\end{equation}
where 
\begin{equation}
 c_1 =-\frac{1}{2\pi}\tr F \ , \qquad  c_2 =\frac{1}{8\pi^2}(\tr F\wedge F-\tr F \wedge \tr F) \ .
\end{equation}

Since ${\rm SU}(2)$ has traceless field strength, then the number $C^2_1$ is zero and $C_2$ simplifies to the usual expression.
While for ${\rm U}(1)$ the inverse happens, i.e., $C_2=0$ and
\begin{equation}
 C^2_1=\frac{1}{4\pi^2}\int\limits_M{F\wedge F} \ .
\end{equation}

So we readily see that for magnetic Reissner-Nordstr\"om $C_1^2=0$. For this configuration all Chern numbers vanish, this points to
a triviality. It does so in the following sense. The configuration is basically the Dirac monopole multiplied trivially with the space $\mathbb{R}^2$.
The fact that the construction is trivial is reflected in the vanishing of the topological numbers.

The manifolds underlying Taub-NUT/Bolt both have a $3-$dimensional boundary, $\mathbb{S}^3_{\infty}$ say. Consequently, the Chern-number may also be calculated
through the Chern-Simons 3-form integrated on this boundary, i.e.
\begin{equation}
 c_1\wedge c_1[M]=cs[\partial M]=\frac{1}{4\pi^2}\int\limits_{\mathbb{S}^3_{\infty}}A\wedge \rd A 
\end{equation}
where we have simplified the Chern-Simons form for the Abelian case.

Before continuing to the next section let us consider a Reissner-Nordstr\"om instanton with electric charge $q$ and 
magnetic charge $p$. Imposing self-duality on the field yields \cite{pope}
\begin{equation}
 F=\frac{p}{r^2}\left(\rd r\wedge \rd\tau-r^2\sin\theta \rd\theta\wedge \rd\phi\right)\ .
\end{equation}
This field has been studied in \cite{etesi} and from there we know that the configuration is characterized by
\begin{equation}
C_1=n \quad \text{and} \quad  C_1^2=2n^2  
\end{equation}
where $n$ is an integer and $2p=n$. Note that self-duality also affects the background geometry. Since the metric parameters comply with $p^2=q^2$ then
the background is in fact Schwarzschild. This is a trivial solution to the Euclidean Einstein-Maxwell equations. Since the energy-momentum tensor vanishes
and so the background is a solution to the vacuum equations.

Recall that the Schwarzschild instanton is asymptotically flat. Moreover, if we consider the same field as described above over a globally flat space, 
it is still a solution to the Euclidean Einstein-Maxwell equations albeit a trivial one. This pair of configurations has the same asymptotic boundary and
the same field content. The integrals involved in the calculation of Chern numbers yield identical results. The key difference is the cohomology
of each background. While it is trivial for flat space it is not for the Schwarzschild instanton. In fact, the second cohomology group
of the Schwarzschild space is infinite cyclic i.e. $\mathbb{Z}$. This means that $2p=n$ only for the Schwarzschild background.

\subsection{Dyonic Taub-Bolt}

In this section, we compare dyonic and self-dual fields over Taub-NUT and Taub-Bolt spaces. To this end, 
we first review some
known results for the Taub-NUT space \cite{pope} some of which carry over to the Taub-Bolt space. Firstly, 
recall that the biaxial symmetry present in the Taub spaces is ${\rm U}(1)$. 
The Killing form associated to this symmetry, $\xi$ say, may be used to define
a ${\rm U}(1)$ two-form $F=\lambda \rd\xi$, where $\lambda$ is related to the electromagnetic charges. In turn,
$F$ has the (anti) self-dual components
\begin{equation}
 F_{\pm}=\frac{\lambda(M\pm N)}{(r\pm N)^2}\left[\rd r\wedge(\rd\tau+2N\cos\theta \rd\phi)\mp(r^2-N^2)\sin\theta \rd\theta\wedge \rd\phi\right]\ .
\end{equation}
This should be compared with the dyon field whose (anti) self-dual components are obtained from the above expression by changing 
\begin{equation}
 \lambda(M\pm N)\rightarrow 1/2(\pm p-q)\ .
\label{mn}
\end{equation}

When $M=N$, the famous  self-dual field strength (on Taub-NUT) is recovered \cite{pope,eh,sen,etesi}
\begin{equation}
 F=\frac{p}{(r+N)^2}\left[\rd r\wedge(\rd\tau+2N\cos\theta \rd\phi)-(r^2-N^2)\sin\theta \rd\theta\wedge \rd\phi\right] \ .
\end{equation}
Since the particular case $M=-N$ is not physical, no anti self-dual field is possible over the Taub-NUT space. 
Additionally, note that
both (anti) self-dual fields comply with $p^2-q^2=0$. This means that the dyon over a Taub-NUT space does not modify the
background. This is a particular distinction between (anti) self-dual fields over Riemannian and Lorentzian backgrounds, 
as pointed out in \cite{ktb}. 
The energy-momentum tensor vanishes for fields with this symmetry and so vacuum solutions may be endowed with them
without further changes. This is to say, they are trivial solutions to the Euclidean Einstein-Maxwell equations.

For the dyon over the Taub-Bolt space this is not so. The dyon field is not (anti) self-dual and so the charged Taub-Bolt geometry 
is different from its neutral counterpart. As mentioned above, consistency requires $V_d(r_+)=0$ and $h(r_+)=0$. 
In this case, $r_+=r_b$, the only physical solution $(r_b>N)$
which satisfies the cubic equation
\begin{equation}
 r_+^3-2N(1+\nu^2)r_+^2+2\nu^2N^3=0
\end{equation}
where $\nu$ is short hand for $p/2N$.

We recall that the asymptotic behavior of Taub-NUT and Taub-Bolt spaces is identical. For example,
the total magnetic flux\footnote{For comparison with Chern invariants we normalize the flux with a 1/2$\pi$ factor.} is
\begin{equation}
 \varPhi=\frac{1}{2\pi}\int\limits_{\mathbb{S}^2_{\infty}} F=-2p \ .
\end{equation}
However, the cohomology of Taub-NUT is trivial and so we may not conclude that this flux is quantized in general. 
This means that all continuous values of $p$ are admissible and since $F$ is self-dual, it corresponds to a continuous
family of instantons with finite action \cite{etesi}.
The situation is completely different in the case of the Taub-Bolt space since its second cohomology group is
infinite cyclic, $H^2(\mathbb{CP}^2\char`\\\{*\})=\mathbb{Z}$.

The Chern number is calculated as
\begin{equation}
 \frac{1}{4\pi^2}\int{F\wedge F}=4p^2
\end{equation}
where the domain of the integral is not indicated because the result is the same for Taub-NUT  ($\mathbb{R}^4$) and Taub-Bolt 
($\mathbb{CP}^2\char`\\\{*\}$). 
This is due to the fact that both dyons vanish at the source (i.e. nut or bolt, depending on the case) and both
have the same value at infinity. An alternative way to see this is to use the fact that both spaces have the same asymptotic boundary and, therefore, the Chern-Simons invariant in both cases is
\begin{equation}
 \frac{1}{4\pi^2}\int\limits_{\mathbb{S}^3_{\infty}}{A\wedge \rd A}=4p^2 \ .
\end{equation}
Nevertheless, this number is an integer only for the Taub-Bolt space. In fact, 
this invariant classifies the different possible gauge potentials --- different according to their
magnetic charge. This may be paraphrased to saying that the magnetic charge is a topological quantum number for Taub-Bolt.
For Taub-NUT, however, the continuous family of fields are referred to as non-topological \cite{etesi}, since  the above quantities are non-zero but are not predictable from the background topology.

As a final remark we mention that for ${\rm U}(1)$ bundles, $\mathbb{S}^2$ backgrounds are of interest ---  as we know from the Dirac monopole. 
Since bundle theory is homotopy invariant\footnote{This
is a corollary of Steenrod's classification theorem  \cite{steenrod}.}, the same applies
for manifolds of higher dimension but with the homotopy type of a two-sphere, e.g., the  Taub-Bolt space.
Integral cohomology can be forced over the NUT space \cite{pope} however, it emerges naturally on the Bolt scenario.
Chern classes have already been studied for (anti)self-dual fields over rotating Taub-Bolt spaces \cite{ktb}.
However, to the best of our knowledge the topological quantum numbers of Taub-Bolt dyons (which are not self-dual) have 
not been studied in the literature before.

\section{Conclusions}
\label{sec:con}

In this work, we have analyzed the gravitational instantons that arise from the Taub-NUT spacetime with electric and magnetic charges. 
In particular, we investigated the topological properties of dyons and (anti) self-dual fields on Euclidean backgrounds corresponding to 
Taub-NUT and Taub-Bolt spaces. We conclude that instantons with a bolt carry topological quantum numbers. Their cohomology is integral
and there are countably infinite distinct ${\rm U}(1)$-bundles over them. We have shown that the Chern numbers of Taub-Bolt dyons are closely 
related to the invariants of the famous Taub-NUT self-dual field. In terms of physical parameters they are the same. The essential difference
is that the former are quantized due to topological reasons.

Taub-NUT configurations are usually considered as the gravitational analog of Dirac's magnetic monopole. Our results show that strictly speaking
this is not true.  Indeed, Dirac's monopole is treated as a quantum system as a result of the underlying space topology. We have shown 
that in gravity this role is played by Taub-Bolt configurations, only. We believe that this result might be important for the quantization 
of gravity by using the path-integral approach. In our opinion, Taub-Bolt instantons should be used to construct the physical vacuum state
of gravity. The consequences of this alternative approach are difficult to predict in advance. We expect to handle this problem in a future 
investigation.

The study of topological defects is also important in the context of topological quantization in which, however, the underlying geometry should be Lorentzian. The results obtained in this work could serve as a guide for  investigating  Lorentzian Taub configurations, which are of particular importance when interpreted as anisotropic cosmological models. The information obtained from the analysis of the topological invariants of such Lorentzian configurations would imply the first step towards the construction of an alternative theory called topological quantum cosmology. This issue will be investigated in future works.

Introducing a cosmological constant to the dyonic instantons does not alter their topology. In fact, our results remain the same if the ALF
configurations we have studied are generalized to be asymptotically locally AdS$_4$. Through the AdS/CFT correspondence
\cite{ads1,ads2,ads3,ads4} dyonic AdS$_4$ black holes
have already been investigated and related to magnetohydrodynamics \cite{mhd} and Hall conductivity \cite{hall}. This points to further exploring the dyonic Taub-NUT/Bolt
system.

\section*{Acknowledgements}
DFA is supported by CONACyT, Grant No. 404449. 
This work has been supported by the UNAM-DGAPA-PAPIIT, Grant No. IN111617.

\label{last}
\end{document}